\documentclass{article}
\usepackage{spconf,amsmath,graphicx,algorithm2e,multirow,tikz,colortbl,balance}
\usetikzlibrary{shapes.geometric}
\usepackage{textcomp}
\usepackage{hyperref}

\definecolor{lightgray}{rgb}{0.9,0.9,0.9}

\usepackage{fancyhdr}
\usepackage{textcomp}
\usepackage[absolute]{textpos}
\newcommand{\copyrightstatement}{
	\begin{textblock}{15}(0.4,0.2)
	\noindent
	\textblockcolour{white}
	\copyright 2024 IEEE. Published in 2024 International Conference on Image Processing (ICIP), scheduled for 27-30 October 2024 in Abu Dhabi, UAE. Personal use of this material is permitted. However, permission to reprint/republish this material for advertising or promotional purposes or for creating new collective works for resale or redistribution to servers or lists, or to reuse any copyrighted component of this work in other works, must be obtained from the IEEE. DOI: \url{10.1109/ICIP51287.2024.10647688}
	\end{textblock}
}

\title{Conditional optimal filter selection for multispectral object classification} 

\name{Katja Kossira, David Sch\"on, J\"urgen Seiler, and Andr\'e Kaup\thanks{The authors gratefully acknowledge that this work has been supported by the Bayerische Forschungsstiftung (BFS, Bavarian Research Foundation) under project number AZ-1547-22}}
\address{Multimedia Communications and Signal Processing\\
Friedrich-Alexander-University Erlangen-N\"urnberg (FAU)\\
Erlangen, Germany\\
\{katja.kossira, david.schoen, juergen.seiler, andre.kaup\} @fau.de}

\begin{document}
\copyrightstatement
%\flushbottom
\RestyleAlgo{ruled}
\maketitle

\begin{abstract}
Capturing images using multispectral camera arrays has gained importance in medical, agricultural and environmental processes. However, using all available spectral bands is infeasible and produces much data, while only a fraction is needed for a given task. Nearby bands may contain similar information, therefore redundant spectral bands should not be considered in the evaluation process to keep complexity and the data load low. In current methods, a restricted and pre-determined number of spectral bands is selected. Our approach improves this procedure by including preset conditions such as noise or the bandwidth of available filters, minimizing spectral redundancy. Furthermore, a minimal filter selection can be conducted, keeping the hardware setup at low costs, while still obtaining all important spectral information. In comparison to the fast binary search filter band selection method, we managed to reduce the amount of misclassified objects of the SMM dataset from 318 to 124 using a random forest classifier.

\end{abstract}


\section{Introduction}
\label{sec:intro}
\vspace{-0.3cm}
In the last decades, the separation of light into single spectral bands using multispectral imaging (MSI) systems has gained interest in many fields. It is based on the fact that materials absorb and reflect light for specific wavelengths, thus providing applications in food processing such as agriculture crop monitoring \cite{CropMonitoring} or medical examinations like subcutaneous vein detection \cite{VeinDetection} as well as the estimation of burn injury degrees \cite{BurnInjury}.

\begin{figure}
	\centering
	\includegraphics[width=0.28\textwidth]{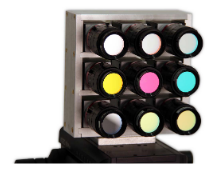}
	\vspace{-0.3cm}
	\caption{A multispectral camera with nine channels \cite{CAMSI}.}
	\label{camsi}
	\vspace{-0.3cm}
\end{figure}

Setups using MSI can vary in their realization. One approach is using multiplexed illumination, where several subsets of filtered LEDs are turned on when recording one image \cite{MultiplexedIllumination}, while another solution introduces a filter wheel in front of a single camera \cite{FilterWheel}. The camera array for multispectral imaging (CAMSI) \cite{CAMSI} shown in Fig. \ref{camsi} is a multicamera multi-filter setup, where each of the nine monochromatic cameras has a filter mounted in front of the camera lens. These filters can be easily replaced by other combinations, depending on the field of use and the region of interest.

Contrary to hyperspectral imaging (HSI), which aims at capturing hundreds of contiguous spectral bands, MSI systems only record at several distinct wavelengths, enabling more cost efficient setups. However, while the filters of an HSI system are implicitly fixed, MSI configurations, especially CAMSI, allow an adaptive choice from a set of available filters. Since this set can include multiple ultraviolet (UV), visible (VIS), and infrared (IR) wavelength filters, the number of possibilities to combine them is very large. 

To evaluate these combinations, several methods have been developed, aiming to approximate the best set of filters in an adequate amount of time. Hardeberg et al. \cite{MOCR} propose to maximize the orthogonality between different filters using a recorded set of the most significant reflectances. Since in specific applications several areas of the spectrum contain characteristics and are therefore much more useful than others, handpicking those filters belonging to wavelength regions which appear to be heuristically interesting is also a common method \cite{HandpickFilters}. Considering that handpicking filters is based on human experience, Sippel et al. \cite{FastBinarySearch} introduced a fast binary search (FBS) algorithm, which is capable of reaching the optimum of a full search while ensuring a constant time for the selection of every desired number of filters. Thus, human errors become negligible.

Even though all of the established methods work well, they are based on finding an optimal filter selection for a fixed number of cameras, disregarding any further constraints. However, for economic and practical reasons a reduction of the number of cameras would be beneficial. By using fewer filters, the required computational power can be reduced, leading to more efficient and resource-friendly imaging systems. Furthermore, it may occur that beyond a certain number of selected wavelengths the classification does not significantly improve. It may even lead to overfitting, where the addition of extra features negatively impacts the multispectral classification performance.

In this paper, the existing FBS approach from \cite{FastBinarySearch} is extended to incorporate specific constraints, resulting in the conditional filter band selection (CFBS) algorithm. While the multispectral classification performance is maximized, the number of filters used is kept as small as possible. Furthermore, the influence of noise can be estimated and limited by making use of this novel approach.

The paper is structured as follows. Section \ref{sec:Related Work} reviews the current state of the art from literature. In Section \ref{sec:Proposed Method} we describe the FBS for an optimal filter selection as well as our novel CFBS approach for conditional optimal filter selection. The results are presented and verified in Section \ref{sec:Verification}, with detailed comparison to conventional solutions. Summary and conclusion are given in Section \ref{sec:Conclusion}.

\vspace{-0.2cm}
\section{Related Work}
\label{sec:Related Work}
\vspace{-0.2cm}
To address the issue of choosing a good filter set for multispectral classification, several algorithms have been developed. 

Hardenberg et al. \cite{MOCR} propose maximizing orthogonality in the characteristic reflectance vector space (MOCR) representative of the application area for the system. The filters are chosen sequentially and aim to capture areas most spectra have in common. Thus, the regions that make the spectra distinguishable are not selected. Li et al. \cite{MLI} developed the maximum linearity independence (MLI) algorithm. In MLI the underlying spectra are not considered. Instead, MLI tries to find a filter configuration such that the filter matrix is well conditioned and therefore the most interesting regions for discrimination cannot be found. 

Employing a uniform search as in \cite{Uniform} is an intuitive, heuristic strategy to choose a set of filters where the dominant wavelengths are relatively equally spaced throughout the complete spectrum. By that the whole wavelength are is covered by the filter selection and no region is left out, considering all available information. However, since only a restricted number of filters can be chosen for setups such as CAMSI \cite{CAMSI}, filters with a high bandwidth have to be chosen for covering the whole spectral range, involving the risk of missing out important information.

A further common method that is approximately intuitive is the full search as employed in \cite{FullSearch}. This approach evaluates all possible filter combinations of the available filters in order to find the best subset of filters for the particular task or problem. Every possible combination of filters from the available pool is considered and evaluated based on some criteria. By that it guarantees to find the best subset of filters. However, with a large filter set to chose from it can be computationally expensive. For example, when selecting 9 out of 150 available filters, 82947113349100 combinations are possible.

To address this problem, Sippel et al. developed the fast binary search \cite{FastBinarySearch}, which is designed to significantly reduce the computational cost compared to the full search, while still providing an optimal filter selection. It is a common method to find a near optimum solution for selecting $N$ out of $K$ filters from an available filter set, while the spectral behavior of the cameras and lenses are assumed to be perfect.

\section{Proposed Conditional Filter Band Selection Algorithm}
\label{sec:Proposed Method}
Comparing the different literature approaches, fast binary search \cite{FastBinarySearch} works best and guarantees a good multispectral classification. Therefore, this approach is extended  and modified further in our work to incorporate specific constraints such as a minimal number of filters to select, or spectral noise restrictions.

The filter matrix \textbf{\textit{F}} of FBS is calculated by 
\begin{equation}
	f_{i,j} = \int_{\lambda_{\min}}^{\lambda_{\max}} c_{i}(\lambda) \cdot l_{j}(\lambda)~d\lambda ~~~~,
	\label{eq:filter matrix}
\end{equation}
where $\lambda_{\min}$ and $\lambda_{\max}$ are the integration bounds determined by the bandpass filter characteristics, $c_{i}(\lambda)$ represents the discretized  filter curve of filter number $i$, and $l_{j}(\lambda)$ is the light spectrum of the $j$-th object to classify. Subsequently, an undirected graph visualizing the distances between the filters is created. The distance metric is the spectral angle \cite{SpectralAngle}, which can be calculated using 
\begin{equation}
	\theta(\textbf{\textit{f}}_{i}, \textbf{\textit{f}}_{j}) = \arccos \left(\frac{\textbf{\textit{f}}_{i}^{T}}{||\textbf{\textit{f}}_{i}||_{2}} \frac{\textbf{\textit{f}}_{j}}{||\textbf{\textit{f}}_{j}||_{2}}\right) ~~~.
	\label{eq:spectral angle}
\end{equation}
Here, $\textit{\textbf{f}}_{i}$ and $\textit{\textbf{f}}_{j}$ denote the $i$-th row and $j$-th line vector respectively. Two filters should not be used together if they contain nearly the same values for all objects to classify, thus resulting in a spectral angle close to zero. The evaluation of the spectral angle for each filter pair combination $a_{i,j} = \theta (\textbf{\textit{f}}_{i},  \textbf{\textit{f}}_{j})$ forms the adjacency matrix \textbf{\textit{A}}. 

In order to maximize the minimal distance within the undirected graph, a binary search over the spectral angle is performed. The current tested spectral angle $\theta_{\text{curr}}$ can be calculated as
\begin{equation}
	\theta_{\text{curr}} = \frac{\theta_{\min} + \theta_{\max}}{2} ~~~,
	\label{eq:current spectral angle}
\end{equation}
where the starting bounds for the binary search are $\theta_{\text{min}} = \text{min(\textbf{\textit{A}})}$ and $\theta_{\text{max}} = \text{max(\textbf{\textit{A}})}$ respectively.
Consequently, the maximum number of connected nodes with the minimal spectral angle $\theta_{\text{min}}$ need to be found in order to know whether a configuration for the current spectral angle $\theta_{\text{curr}}$ exists for which enough filters can be picked. A mixed integer maximization problem \cite{MIP} is set up, whose objective is to maximize the number of nodes that can be chosen. Constraints are added to the optimization problem to prevent picking filters that are connected by an edge with lower edge weight than the current spectral angle
\begin{equation}
	\begin{split}
    \textbf{\textit{s}} &= \underset{\textbf{\textit{s}}}{\text{argmax}} \sum_{i=1}^{K} s_{i} \\
    &~~~~\text{s.t.} \quad s_{i} + s_{j} \leq 1 \quad \forall i \neq j, \phantom{\text{s.t.}} a_{i,j} < \theta_{\text{curr}} ~~~,
	\end{split}
	\label{eq:mip}
\end{equation}        
where $\textbf{\textit{s}}$ is the binary selection vector. In case the resulting sum is lower than $K$, the upper bound of the binary search is set to $\theta_{\text{curr}}$. If the resulting sum is greater or equal to the number of desired filters, the lower bound is set to $\theta_{\text{curr}}$
\begin{equation}
\hspace{-0.5cm}
	\begin{cases}
	\theta_{\min} = \theta_{\text{curr}} & \text{if}~~ \sum^{K}_{i=1} s_{i} \geq N \\
	\theta_{\max} = \theta_{\text{curr}} & \, \text{else}.
	\end{cases}
	\label{eq:bounds}
\end{equation} 
The optimal combination of $N$ out of $K$ available filters is obtained and will be used for calculating the CFBS as described in the next section.

\begin{algorithm}[t!]
	\caption{The basic procedure of CFBS.}\label{alg:CFBS}
	\KwIn{$\text{SNR}_{th}$ = 0,...,30; $\text{cvs}_{th} = 0.90,...,0.98$}
	\KwOut{Minimal filter selection $\textbf{\textit{s}}_{min}$}
	Calculate $\text{SNR}_{k}$ of all filter signals in $\textbf{\textit{s}}_{k}$ using \eqref{eq:SNR};\\
%	$\textbf{\textit{s}}_{curr}$ = [~];\\
	\For{$k < K$}{
		\If{$\mathrm{SNR}_{k} < \mathrm{SNR}_{th}$}{
			Add $\textbf{\textit{s}}_{k}$ to $\textbf{\textit{s}}_{curr}$;\\ 
		}
	}
	Find optimal filter selection \textbf{\textit{s}} using FBS using (\ref{eq:filter matrix}) - (\ref{eq:mip});\\
	Objective: minimize $\textbf{\textit{s}}_{curr}$;\\
	Calculate cross-validation score $\text{cvs}$: classifier(objective, $\textbf{\textit{s}}_{curr}$);\\
	\ForEach{$\textbf{s}_{i} \in \textbf{s}_{curr}$}{
		\eIf{$\mathrm{cvs}_{i} = 1.0$}{
			$\textbf{\textit{s}}_{min} = \textbf{\textit{s}}_{i}$;\\
			$\mathit{\text{cvs}_{k}}$ = 1.0;\\
		}
		{\If{$\mathrm{cvs}_{min} < \mathrm{cvs}_{k}$}
			{$\mathit{\text{cvs}_{min}} = \mathit{\text{cvs}_{i}}$;\\
				$\textbf{\textit{s}}_{min} = \textbf{\textit{s}}_{i}$;\\
			}
		}
	}
\end{algorithm}
\begin{figure}[t]
	\centering
	\begin{tikzpicture}
		\node[anchor=south west, inner sep=0] (image) at (0,0) {\includegraphics[width=0.5\textwidth, height=0.4\textheight, keepaspectratio]{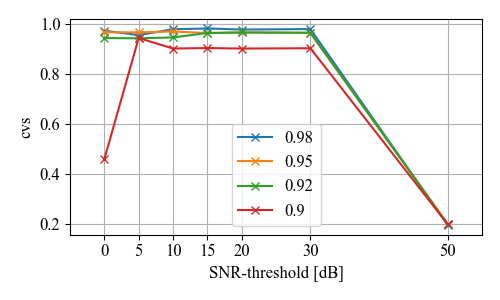}};
		\node[fill=white, minimum height=0.5cm, minimum width=1cm] at (6.1,0.55) {};
	\end{tikzpicture}
	\vspace{-0.9cm}
	\caption{$\text{cvs}$ performance for different combinations of  $\text{cvs}_{th}$ and $\text{SNR}_{th}$.}
	\label{fig:cvs}
	\vspace{-0.5cm}
\end{figure} 

So far when performing a filter selection, typically a fixed number of filters to select is given as a constraint. However, sometimes it is advantageous to chose a minimum number of filters or to add further restrictions and conditions. Therefore, the result of the FBS algorithm will be further processed in the conditional filter band selection approach. Our CFBS method performs a systematic search over the optimal filter selection result $\textbf{\textit{s}}$ and returns the minimum amount $\textbf{\textit{s}}_{\min}$ of filters needed, while still guaranteeing a good multispectral classification of the material
\begin{equation}
\vspace{-0.2cm}
	\textbf{\textit{s}}_{\min} = \text{CFBS} (\textbf{\textit{s}})~~~.
	\label{eq:MinimizationFilterSelection}
\end{equation}
As it is proven that spectral band noise significantly impacts the performance of MSI classification systems \cite{ImpactSpectralNoise}, spectral bands with increased noise levels should not be further considered in the selection process. Since those bands are still included and can be selected during the optimal FBS process, the Signal-to-noise ratio (SNR) \cite{SNR} $\text{SNR}_{k}$ of each of the $K$ spectral filter signal responses $S_{k}$ can be estimated as the the deviation from the mean. This is a common method in image processing, since the power of the signal is unknown and therefore cannot be used. The $\text{SNR}_{k}$ is calculated by
\begin{equation}
	\text{SNR}_{k} = \frac{\mu_{k}}{\sigma_{k}} = \frac{\frac{1}{M} \sum_{i = 1}^{M} x_{i,k}}{\sqrt{\frac{1}{M} \sum_{i = 1}^{M} (x_{i,k} - \mu_{k})^{2}}} ~~~,
	\label{eq:SNR}
\end{equation}
where $\mu_{k}$ represents the mean and $\sigma_{k}$ denotes the standard deviation of filter band $k$. $x_{i}$ corresponds to the single pixel values while $M$ is the number of pixels in the signal region. As soon as the $\text{SNR}_{k}$ falls below a certain threshold $\text{SNR}_{th}$, the respective band is removed from all available wavelengths, considering  that the information loss caused by the noise yields no significant improvement for the multispectral classification process. After the filter band restriction, the FBS is performed, resulting in the optimal filter selection $\textbf{\textit{\^s}}$. Two of these filters are chosen and rated regarding their classification performance and prediction accuracy, which are evaluated using the cross-validation score ($\text{cvs}$) \cite{cvs}. Cross validation is defined as the process of data partitioning to estimate prediction performances. The method involves dividing the available data into pairs of training and test datasets, where the statistical model is fitted to the training data. Mathematically, the cross-validation score can be calculated using 
\begin{equation}
\vspace{-0.2cm}
	\text{cvs} = \frac{1}{k} \sum^{k}_{i=1}\text{Acc}_{i}  ~~~,
	\vspace{-0.1cm}
	\label{eq:cvs}
\end{equation}
where $k$ represents the number of folds and $\text{Acc}_{i}$ the accuracy of the $i$-th fold. Subsequently, the model is evaluated based on the predictions for the test data \cite{cvsProc}. By repeatedly applying this process to different data splits, the average predictive performance of the classifiers is estimated. In case the cross-validation score $\text{cvs}$ reaches a specific cross-validation threshold $\text{cvs}_{th}$, the algorithm terminates by returning the conditional optimal filter selection $\textbf{\textit{s}}_{min}$. In case the threshold is not reached, an additional filter of \textbf{\textit{s}} is chosen and evaluated with $\textbf{\textit{s}}_{min}$ and a new cross-validation score is calculated. This described process is performed iteratively, until no further improvement of the cross-validation score can be achieved. The CFBS procedure is summarized in Algorithm \ref{alg:CFBS}.

The results of different combinations of $\text{SNR}_{th}$ and $\text{cvs}_{th}$ are shown in Fig. \ref{fig:cvs}. We can infer that a $\text{cvs}_{th} < 0.90$ does not yield any additional improvements in the results, therefore $\text{cvs}_{th}$ values below 0.90 are not recommended. Moreover, a $\text{cvs} > 0.98$ could hardly be reached, so we can conclude that a $\text{cvs}_{th} = 0.90,...,0.98$ is the best choice for CFBS. Apart from this, the optimal choice for the $\text{SNR}$-threshold is $\text{SNR}_{th} = 0,...,30$. For higher $\text{SNR}_{th}$-values, only a few wavelengths to chose from are left. Due to a high amount of noise in high wavelength regions, the remaining bands are typically all located in the lower wavelength area, therefore much redundant information has to be included in the filter selection. Table \ref{tab:ResultsCFBS} supports these results.

\textbf{\begin{table}[t]
		\small
		\caption{The amount of filters $n$ and the amount of wrongly classified objects $\text{WCO}$ achieved using the CFBS solution with different $\text{SNR}_{th}$ and $\text{cvs}_{th}$.} 
		\vspace{0.2cm}
		\begin{tabular}{c|c|ccccccc}
			\multicolumn{2}{c}{}& \multicolumn{7}{|c}{$\text{SNR}_{th}$} \\ \cline{3-9}
			\multicolumn{2}{r|}{$\text{cvs}_{th}$} & 0 & 5 & 10 & 15 &  20 & 30 & 50 \\   
			\hline
			\hline
			\multirow{4}{*}{$n$}& 0.98 & 8 & 8 & 8 & 8 & 8 & 7 & 8 \\
			& 0.95 & 6 & 6 & 5 & 4 & 4 & 4 & 9 \\
			& 0.92 & 5 & 5 & 4 & 4 & 4 & 4 & 7 \\
			& 0.90 & 5 & 5 & 3 & 3 & 3 & 3 & 7 \\
			\hline
			\multirow{4}{*}{$\text{WCO}$}& 0.98 & 125 & 197 & 110 & 104 & 112 & 111 & 3990 \\
			& 0.95 & 169 & 178 & 165 & 182 & 176 & 171  & 4015 \\
			& 0.92 & 289 & 278 & 271 & 190 & 174 & 182 & 4020 \\
			& 0.90 & 286 & 279 & 489 & 491 & 486 & 486 & 4011 \\
		\end{tabular}
		\label{tab:ResultsCFBS}
\end{table}}
              \vspace{-0.5cm}     
\section{Verification}
\label{sec:Verification}

\begin{figure}[t]
	\centering
	\begin{tikzpicture}[auto]
		\node (dataset) [draw, fill=gray, minimum height=0.7cm, minimum width=7cm, align=center] {Dataset};		
				
		\node (11o5) [below of=dataset, draw, fill=lightgray,  xshift=-2.95cm, yshift=0.2cm] {Part 1};
		\node (12o5) [below of=dataset, draw, fill=white,  xshift=-1.5cm, yshift=0.2cm] {Part 2};
		\node (13o5) [below of=dataset, draw, fill=white, yshift=0.2cm] {Part 3};
		\node (14o5) [below of=dataset, draw, fill=white,  xshift=1.5cm, yshift=0.2cm] {Part 4};
		\node (15o5) [below of=dataset, draw, fill=white,  xshift=2.95cm, yshift=0.2cm] {Part 5};
		\node (1) [left of=11o5] {1};
		
		\node (21o5) [below of=11o5, draw, fill=white, yshift=0.3cm] {Part 1};
		\node (22o5) [below of=12o5, draw, fill=lightgray, yshift=0.3cm] {Part 2};
		\node (23o5) [below of=13o5, draw, fill=white, yshift=0.3cm] {Part 3};
		\node (24o5) [below of=14o5, draw, fill=white, yshift=0.3cm] {Part 4};
		\node (25o5) [below of=15o5, draw, fill=white, yshift=0.3cm] {Part 5};
		\node (2) [left of=21o5] {2};

		\node (31o5) [below of=21o5, draw, fill=white, yshift=0.3cm] {Part 1};
		\node (32o5) [below of=22o5, draw, fill=white, yshift=0.3cm] {Part 2};
		\node (33o5) [below of=23o5, draw, fill=lightgray, yshift=0.3cm] {Part 3};
		\node (34o5) [below of=24o5, draw, fill=white,yshift=0.3cm] {Part 4};
		\node (35o5) [below of=25o5, draw, fill=white, yshift=0.3cm] {Part 5};
		\node (3) [left of=31o5] {3};
		
		\node (41o5) [below of=31o5, draw, fill=white, yshift=0.3cm] {Part 1};
		\node (42o5) [below of=32o5, draw, fill=white,yshift=0.3cm] {Part 2};
		\node (43o5) [below of=33o5, draw, fill=white,yshift=0.3cm] {Part 3};
		\node (44o5) [below of=34o5, draw, fill=lightgray,yshift=0.3cm] {Part 4};
		\node (45o5) [below of=35o5, draw, fill=white,yshift=0.3cm] {Part 5};
		\node (4) [left of=41o5] {4};
		
		\node (51o5) [below of=41o5, draw, fill=white,yshift=0.3cm] {Part 1};
		\node (52o5) [below of=42o5, draw, fill=white,yshift=0.3cm] {Part 2};
		\node (53o5) [below of=43o5, draw, fill=white,yshift=0.3cm] {Part 3};
		\node (54o5) [below of=44o5, draw, fill=white,yshift=0.3cm] {Part 4};
		\node (55o5) [below of=45o5, draw, fill=lightgray,yshift=0.3cm] {Part 5};
		\node (5) [left of=51o5] {5};
		
	\end{tikzpicture}
	\caption{Data partitioning for the computation of the cross-validation score. Lightgray highlighted boxes represent the test data, white boxes indicate the training data.}
	\label{fig:cvsProcedure}
\end{figure}
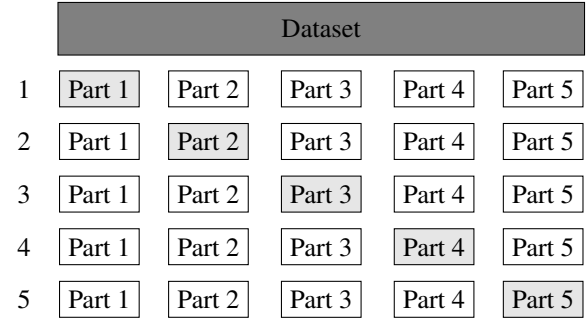
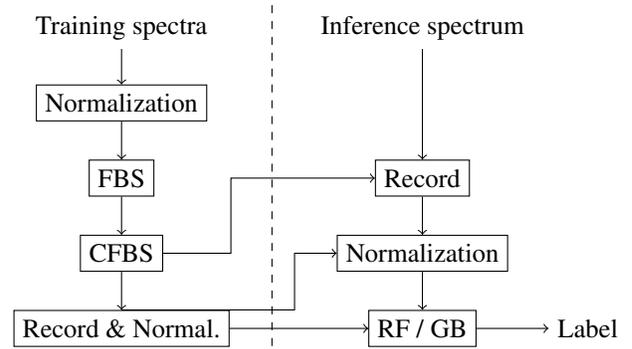
\begin{figure}
	\begin{tikzpicture}
		\node(training) [align=center] {Training spectra};
		\node(normTrain) [below of=training, draw, fill=white] {Normalization};
		\node(FBS) [below of=normTrain, draw, fill=white] {FBS};
		\node(CFBS) [below of=FBS, draw, fill=white] {CFBS};
		\node(recNorm) [below of=CFBS, draw, fill=white] {Record \& Normal.};
		
		\node(inference) [right of=training, xshift=3cm] {Inference spectrum};
		\node(record) [below of=inference, draw, fill=white, yshift=-1cm] {Record};
		\node(normInf) [below of=record, draw, fill=white] {Normalization};
		\node(GradientRandom) [below of=normInf, draw, fill=white, align=center] {RF / GB};
	
		\node(label) [right of=GradientRandom, xshift=1.2cm] {Label};
		
		\draw[->] (training) -- (normTrain);
		\draw[->] (normTrain) -- (FBS);
		\draw[->] (FBS) -- (CFBS);
		\draw[->] (CFBS) -- (recNorm);
		\draw[->] (inference) -- (record);
		\draw[->] (record) -- (normInf);
		\draw[->] (normInf) -- (GradientRandom);
		\draw[->] (recNorm) -- (GradientRandom);
		\draw[->] (GradientRandom) -- (label);
		\draw[->] (CFBS.east) --++ (0.9,0) |- (record.west);
		\draw[->] (recNorm.north) --++ (2.3,0) |- (normInf.west);
		\draw[dashed, overlay] (2,0.3) -- (2,-4.3);
				
	\end{tikzpicture}
	\caption{The filter selection and multispectral classification procedure of CFBS.}
	\vspace{-0.4cm}
	\label{fig:trainingInference}
\end{figure}

This paper is the first to propose a minimal filter selection, therefore a comparison to similar approaches is not possible. Instead, in order to verify the quality of the filter selection, the experiment is divided into two sections. On the one hand, our method is evaluated focusing solely on the multispectral classification accuracy for different $\text{SNR}$- and $\text{cvs}$-thresholds, resulting in the minimal filter selection number $\textbf{\textit{s}}_{min}$ accordingly. On the other hand, the output is compared to the FBS \cite{FastBinarySearch} and other approaches from literature with regard to the multispectral classification accuracy and number of chosen filters, as well as the number of wrongly classified objects. 

Both parts require an evaluation dataset. Since the FBS was tested and evaluated on the SMM dataset introduced in \cite{DataBaseFBS}, and a comparison between the FBS and CFBS algorithm shall be performed, the same dataset is also chosen for this paper. It contains spectral measurements of 50 different objects with each of them providing 100 spectra. The objects are categorized in five classes: metal, plastic, wood, paper and fabric. The set of filters used for testing consists of 40 bandpass filters including two different bandwidths (10 nm and 50 nm) in the range from 316 nm to 791 nm. 

To test the actual capability of the CFBS approach, the wavelengths with a high $\text{SNR}$ value above a given $\text{SNR}$ threshold $\text{SNR}_{th}$ are excluded and cannot be chosen during the FBS procedure, since those spectral bands do not contribute sufficient information. The FBS is performed and 9 out of 287 available wavelength regions are selected in order to simulate the CAMSI setup \cite{CAMSI}. The bandwidths of the different filters are set to 10 nm and 50 nm. CFBS starts with using two of the resulting filters of the optimal filter choice \textit{\textbf{s}}, with which we try to classify the objects using the Random Forest \cite{RandomForest} or Gradient Boosting \cite{GradientBoosting} algorithm. The former creates independent decision trees where each of them is trained on a subset of the data and a random subset of features. The prediction is determined by a majority vote or an average prediction across all trees. Since a random selection of subsets is used in the training process, the method is more robust to overfitting. Gradient Boosting on the other hand, creates the decision trees sequentially, meaning that each tree is built in order to correct the errors of the previous tree. Therefore, it is not so robust against overfitting.
\begin{figure}
		\centering
			\begin{tikzpicture}
				\node[anchor=south west, inner sep=0] (image) at (0,0) {\includegraphics[width=0.5\textwidth, height=0.4\textheight, keepaspectratio]{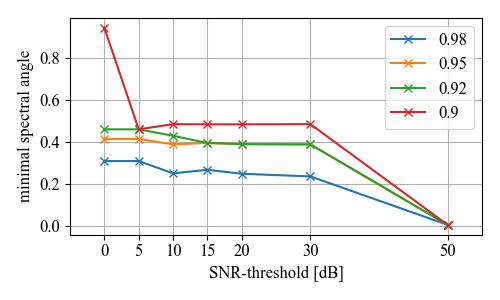}};
				\node[fill=white, minimum height=0.5cm, minimum width=1cm] at (6.1,0.55) {};
			\end{tikzpicture}
		\vspace{-0.9cm}
		\caption{$\theta_{min}$ for different combinations of $\text{cvs}_{th}$ and $\text{SNR}_{th}$}
		\label{fig:minspecang}
		\vspace{-0.2cm}
\end{figure}
\begin{table}[t]
	\centering
	\vspace{-0.4cm}
	\caption{Average number of wrongly classified objects using the proposed CFBS with $\text{cvs}_{th} = 0.98, 0.95, 0.92, 0.90$.}
	\vspace{0.2cm}
	\begin{tabular}{c|c|c|c}
		 \textbf{$\text{CFBS}_{0.98}$} & \textbf{$\text{CFBS}_{0.95}$} & \textbf{$\text{CFBS}_{0.92}$} & \textbf{$\text{CFBS}_{0.90}$} \\
		\hline
		124 & 174 & 222 & 429
	\end{tabular}
	\label{tab:wrongClassification}
	\vspace{-0.3cm}
\end{table} 

\begin{table*}[t]
	\centering
	\caption{Average number of wrongly classified objects using filter selection methods from literature ($n = 9$), uniformly placed filters ($n = 9$), filters chosen by FBS ($n = 9$), and the proposed CFBS ($n_{0.98}= 8$, $n_{0.95} = 4$) with $\text{cvs}_{th} = 0.98, 0.95.$}
	\vspace{0.2cm}
	\begin{tabular}{c|c|c|c|c||c|c}
		\textbf{MOCR} \cite{MOCR} & \textbf{MLI} \cite{MLI} & \textbf{Uniform} \cite{Uniform} & \textbf{Full Search} \cite{FullSearch} & \textbf{FBS} \cite{FastBinarySearch} & \textbf{$\text{CFBS}_{0.98}$} & \textbf{$\text{CFBS}_{0.95}$} \\
		\hline
		459 & 388 & 350 & 320 & 318 & 124 & 174 
	\end{tabular}
	\label{tab:wrongClassificationComp}
\end{table*}

In this method, the available data is divided into $k$ equally sized subsets, resulting in $k$ different pairs of training and test datasets. Given that each dataset comprises a substantial number of 5000 measurements, we use $k = 5$. The data partitioning is illustrated in Fig. \ref{fig:cvsProcedure}, while the block diagram for the training and inference procedure is shown in Fig. \ref{fig:trainingInference}. We can infer that the data is divided into five equally sized subsets. This leads to a total of five computation runs, where each subset serves as the test dataset while the other four subsets are used for training. As a result, five separate values for prediction accuracy are obtained. These values are then aggregated to obtain an average, allowing for a robust estimation of the overall model accuracy. Our tests are based on the $k$-fold cross-validation \cite{K-Fold}, which is a conventional form of data splitting. The training spectra are normalized to ensure the consistency and comparability of the spectral data across all objects. FBS followed by CFBS are executed to find the minimal optimal filter selection for the given dataset. Afterwards, normalization parameters and normalized database vectors are calculated, to which the inference vector is tested against. This way, the label can be estimated. The first step in the multispectral classification procedure is to record the spectrum using the minimal optimal filter selection from CFBS. Afterwards, the normalization parameters from the training step are applied to the recorded multispectral bands. This is necessary, since higher values that may result from a higher filter bandwidth would result in a higher influence on the spectral angle. The Random Forest (RF) and Gradient Boosting (GB) classifiers are used to classify the spectrum of interest, followed by the calculation of the cross-validation score in order to check the accuracy of our proposed method. 
\begin{figure}[t]
	\begin{minipage}[b]{0.49\textwidth}
		\centering
		\includegraphics[width=0.67\textwidth]{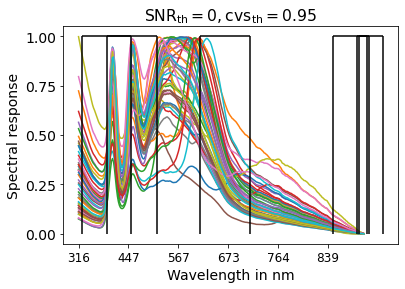}
		\label{fig:subfigure4-a}
	\end{minipage}
	\begin{minipage}[b]{0.49\textwidth}
		\centering
		\includegraphics[width=0.67\textwidth]{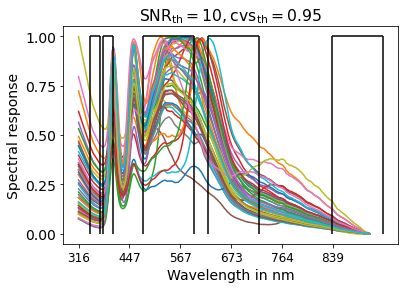}
		\label{fig:subfigure4-b}
	\end{minipage}
	\begin{minipage}[b]{0.49\textwidth}
		\centering 
		\includegraphics[width=0.67\textwidth]{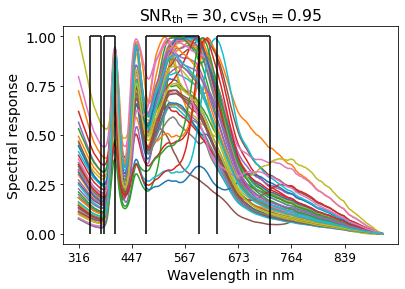}
		\label{fig:subfigure4-c}
	\end{minipage}
	\vspace{-0.7cm}
	\caption{All spectra of different objects that shall be separated visualized by different colors and the result of CFBS in black. $\text{cvs}_{th}= 0.95$ was chosen for all tests. $\text{SNR}_{th} = 0$ at the top, $\text{SNR}_{th} = 10$ in the middle and $\text{SNR}_{th} = 30$ at the bottom.}
	\label{fig:spectraCFBS}
	\vspace{-0.3cm}
\end{figure}

The spectral results of the CFBS with $\text{SNR}_{th} = 0, 10, 30$ and $\text{cvs}_{th} = 0.95$ are shown in Fig. \ref{fig:spectraCFBS}. For $\text{SNR}_{th} = 30$, no high wavelength areas in the selection can be found, since those were excluded by the noise restriction condition, while $\text{SNR}_{th} = 0$ still includes 3 filters that are located in the infrared range. Interesting regions to classify the different spectra are covered. The result may not be very intuitive, as humans typically position filters in areas where the spectral order undergoes changes. However, the highest spectral angle often occurs in regions characterized by maximum changes, which are typically flat. In order to avoid overlaps, and thus redundant information in the covered spectrum, the filters that are located close to each other are chosen to cover 10nm, while filters with high distances cover a higher spectrum and therefore are selected as 50nm. %Furthermore, during the FBS procedure, the spectral angle normalizes the recorded value of a filter first.

In Table \ref{tab:ResultsCFBS}, the results of the amount of selected filters $n$ and the wrongly classified objects WCO with regard to different $\text{cvs}$- and SNR-thresholds are listed. We can infer that significant improvements with regard to the FBS represented by a smaller amount of selected filters ($n < 9$) and wrongly classified objects (WCO $<$ 318) could be achieved. In case the $\text{SNR}_{th}$ is set to a value between 0 and 30 and $\text{cvs}_{th} = 0.90,...0.98$, $n$ and WCO are notably reduced. With an increasing $\text{SNR}_{th}$, $n$ and WCO decrease. This is to be expected since the noise within the spectral bands decreases and consequently more information is available. The best result in terms of wrongly classified objects can be achieved using $\text{cvs}_{th} = 0.98$ and $\text{SNR}_{th} = 15$. But this comes with a required number of filters of still $n = 8$. When using $n = 4$ filters, $\text{cvs}_{th} = 0.95$ and $\text{SNR}_{th} = 30$, a very good multispectral classification with a low number of wrongly identified objects can also be achieved, which is why this combination should be preferred. Nevertheless, the number of wrongly classified objects also remains more or less constant for different $\text{SNR}_{th}$, but varies with $\text{cvs}_{th}$. For $\text{cvs}_{th} = 0.90$ in general, the amount of misclassifications already increases. From Fig. \ref{fig:minspecang} we can infer that the minimal spectral angle also remains more or less constant for different $\text{cvs}_{th}$ and $\text{SNR}_{th}$, but compared with the amount of filters $n$ it is obvious, that it is decreasing  for a larger number of selected filters. This behaviour is reasonable, since an increased number of optimized points in a 2D-plane results in decreasing angles. 

Table \ref{tab:wrongClassification} shows the incorrectly classified objects on average with an $\text{cvs}_{th} = 0.98, 0.95, 0.92, 0.90$ using our approach. We can infer that our method works best with an $\text{cvs}_{th}$ of 0.98. However, this comes with an increased number of selected filters of 8. With the given task of minimizing the total of selected filters, a better solution can be achieved with $\text{cvs}_{th} = 0.95$ or 0.92, where we still obtain a good spectral classification and a significantly reduced number of wrongly classified objects with $n = 4$ or 5.

Table \ref{tab:wrongClassificationComp} compares the performance of our proposed CFBS algorithm with the approaches from literature presented in Section \ref{sec:Related Work}. To this end, the incorrectly classified objects averaged for the different approaches are shown. The CFBS algorithm reduces the number of misclassifications compared to the FBS, which has been the best performing method so far, by over 60~\% when using $\text{cvs}_{th} = 0.98$. Further, the CFBS outperforms the FBS by 45~\% while only using $n = 4$ filters with $\text{cvs}_{th} = 0.95$. With the given task of minimizing the amount of required filters while still misidentifying objects as little as possible, $\text{cvs}_{th}$ = 0.95 should be chosen.

The difference in performance is expected, since MOCR \cite{MOCR} tries  to solve the problem of finding filters that capture areas most spectra have in common. However, we want to filter out the regions that make the spectra distinguishable. In MLI \cite{MLI} the underlying spectra are not considered. Instead, MLI tries to find a filter configuration such that the filter matrix is well conditioned and therefore the most interesting regions for discrimination cannot be found. Furthermore, the approaches from literature include filter bands with a high amount of noise, which do not contribute reliable information. CFBS on the other hand excludes those bands, therefore they are not included in the training process. Instead, bands with less redundant information are chosen and even a lower amount of filters is sufficient to already ensure a good multispectral classification. This means that it is not necessary to always select a fixed number of filters, but gives the opportunity to include further conditions such as noise restrictions or different filter bandwidths.

\balance
\flushbottom
\vspace{-0.2cm}
\section{Conclusion}
\label{sec:Conclusion}
\vspace{-0.2cm}
This paper introduced a novel method to chose a limited number of filters based on previously selected constraints using the conditional filter band selection. It was shown that for all cases our approach reaches a minimum number, while still guaranteeing a high multispectral classification accuracy. Even though the underlying fast binary search finds an optimal solution, FBS does not encounter restrictions, such as noisy spectral bands or the request of keeping the hardware costs and complexity low for an industrial use. In contrast, our conditional filter band selection algorithm achieves to incorporate these restrictions and even improves the multispectral classification accuracy. The number of wrongly classified objects could be significantly reduced, compared to state-of-the-art approaches from literature. The experiment also reveals a good performance for a real-world classification application using the set of filters chosen by our conditional filter band selection algorithm. Future work will consider a 2-layer conditional optimal filter selection model, classifying the objects by their material groups first, followed by the identification of the object itself.

\vfill
\pagebreak
\balance
\flushbottom

\bibliographystyle{IEEEbib}
\bibliography{strings}

\end{document}